%% file: gain_calib_sup_mnras.tex
\documentclass[a4paper,fleqn,usenatbib,useAMS]{mnras}

\usepackage{amsmath,amssymb,amsfonts}
\usepackage{algorithmic}
\usepackage{graphicx}
\usepackage{epstopdf}
\usepackage{enumerate}
\usepackage{subfig}
\usepackage{epsfig}
\usepackage{float}
\usepackage{xcolor}

\include{notations}

\usepackage[T1]{fontenc}
\usepackage{ae,aecompl}

\usepackage{newtxtext,newtxmath}
\title[Signal suppression due to Calibration]{Quantifying Suppression of the Cosmological 21-cm Signal due to Direction Dependent Gain Calibration in Radio Interferometers}

\author[A. Mouri Sardarabadi \& L. V. E. Koopmans]{A. Mouri Sardarabadi$^{1}$ \& L. V. E. Koopmans$^{1}$
\\
\\
$^{1}$Kapteyn Astronomical Institute, University of Groningen, P.O. Box 800, 9700 AV Groningen, The Netherlands}

\date{Last updated 2018 Aug 09; in original form 2018 August 9}

\pubyear{2018}

\begin{document}
\label{firstpage}
\pagerange{\pageref{firstpage}--\pageref{lastpage}}
\maketitle

\begin{abstract}
The 21-cm signal of neutral hydrogen -- emitted during the Epoch of Reionization -- promises to be an important source of information for the study of the infant universe. However, its detection is impossible without sufficient mitigation of other strong signals in the data, which requires an accurate knowledge of the instrument. Using the result of instrument calibration, a large part of the contaminating signals are removed and the resulting residual data is further analyzed in order to detect the 21-cm signal. Direction dependent calibration (DDC) can strongly affect the 21-cm signal, however, its effect has not been precisely quantified. 

In the analysis presented here we show how to exactly calculate what part of the 21-cm signal is removed as a result of the DDC. We also show how a-priori information about the frequency behavior of the instrument can be used to reduce signal suppression. The theoretical results are tested using a realistic simulation based on the LOFAR setup. Our results show that low-order smooth gain functions (e.g.\ polynomials) over a bandwidth of $\sim$10\,MHz -- over which the signal is expected to be stationary -- is sufficient to allow for calibration with limited, quantifiable, signal suppression in its power spectrum. We also show mathematically and in simulations that more incomplete  sky models  lead to larger 21-cm signal suppression, even if the gain models are enforced to be fully smooth. This result has immediate consequences for current and future radio telescopes with non-identical station beams, where DDC might be necessary (e.g.\ SKA-low).  
\end{abstract}


%%%%%%%%%%%%%%%%%%%%%%%%%%%%%%%%%%%%%%%%%%%%%%%%%%%%%%%%%%%%%%%%%%

\section{Introduction}

Studying the redhifted 21-cm hyperfine transition line of neutral hydrogen (called the "21-cm signal" hereafter) originated from the infant Universe is one the most powerful methods for gaining insight into the evolution of our Universe \citep[see e.g.][for reviews]{Furlanetto:2006bq, Morales:2009p1425, 2012RPPh...75h6901P}. The signals generated at a radio receiver in response to these radiations are very weak and their detection is a challenging task which requires highly sensitive instruments \citep[e.g.][]{2006ApJ...653..815M, furlanetto2006181, Parsons:2012dfa, 2013MNRAS.429L...5B, harker2010power, 2015aska.confE...1K, DeBoer:2017ix}. Modern radio telescopes, such the LOw Frequency ARray (LOFAR) in the Netherlands, extend over large distances and combine many receivers to boost both their sensitivity (i.e. collecting area) and angular resolution \citep[see][for details]{lofar_ref}. However, the signal of interest is not the only source of radiation. There are many natural radiators, such as our own Milky Way and other (radio) galaxies, which also generate detectable signals at the receivers, often exceeding the 21-cm signal by many orders of magnitude. Separating these different (partly polarized) signals is an essential part of the detection process \citep[e.g.][]{2002ApJ...564..576D, 2003MNRAS.346..871O, 2008MNRAS.389.1319J, 2015MNRAS.451.3709A, mertens2018statistical}.
During the observation with a radio telescope, knowledge of the known radio sources, a.k.a. the ``sky model", is often used to estimate the unknown (time and frequency dependent) parameters of the telescope which can affect the measurements. This process is called "calibration" and will be discussed in more details in Sec.~\ref{sec:calib}. Calibration is an important part of the data processing of a radio telescope and has been extensively studied in the context of aperture arrays such as LOFAR \citep[e.g.][]{aj02calibration,wijnholds2009,kazemi01102013,yatawatta2015distributed,smirnov2015radio,mouri2014}. 
As the resolution and sensitivity of the instruments increases, more foreground sources can be detected, and often many of the fainter or more diffuse sources which are not included in the sky model during the calibration process. This is the called "sky- or model- incompleteness" and can lead to several artifacts in the final data \citep[e.g.][]{Wijnholds:2016ge, grobler2014calibration, Grobler:2016kp, 2016MNRAS.463.4317P, barry2016calibration, 2017MNRAS.470.1849E}. It is therefore important to make a distinction between the known and unknown foreground sources. The 21-cm signal and often also the extended diffuse Galactic foreground (even if known) are not part of the sky model either because they are simply too faint or too complex to be described by a limited set of model parameters.
In our analysis, we are interested in (possible) corruption of the desired signal, in this case the weak 21-cm signal, as a result of the process of calibration. We also study the effect of sky-incompleteness and the frequency behavior of the instrument. However, the mathematical framework introduced in this analysis is not limited to 21-cm signal studies and can easily be extended to any signal of interest. Our approach also extends previous analyses to that of direction dependent calibration, which is likely necessary for instruments for non-identical receiver beams. 
In Sec.~\ref{sec:data_model}, we discuss the commonly used signal processing data model or measurement equation and our assumptions. Based on this model we formulate calibration as a least squares optimization problem in Sec.~\ref{sec:calib} and then proceed to study the effects of such calibration on the desired signals. In Sec.~\ref{sec:sim}, we use parameters from the LOFAR telescope to study the effect of calibration by comparing the spectra before and after calibration. In Sec.\ref{sect:concl} we draw our conclusions.

%%%%%%%%%%%%%%%%%%%%%%%%%%%%%%%%%%%%%%%%%%%%%%%%%%%%%%%%%%%%%%%%%%

\section{The Radio Telescope Data Model}\label{sec:data_model}

In this analysis we assume to have access to sampled output of $P$ receivers (together called an "array") which are exposed to electromagnetic radiation from extra-terrestrial radio sources. It is possible for each receiver to be a beam-formed array itself, and the model presented here is adequate for any general array topology  \citep[see for example the hierarchical structure of the LOFAR telescope;][]{lofar_ref}. As the result of Earth rotation, the apparent position of the sources change in time, as does the array with respect to a reference point on the sky. 
We assume that the sources are stationary (i.e.\ not changing in brightness nor structure) during the observation of $N$ samples of the electromagnetic signal/voltages in time.
%, where $N$ depends on the angular resolution of the instrument. 
A single observation with $N$ samples is denoted as a "snapshot" in time. We also assume that the output is divided into $K$ frequency channels for which the narrow-band assumption hold \citep[e.g.][]{wijnholds2009, mouri2016PhD}. 
We used the same sky model given by \citet{yatawatta2015distributed}, where the sources on the sky are grouped spatially into so-called clusters. We assume that all sources within a cluster are affected by the same complex and direction dependent gain. In this work we assume that the sky is unpolarized. The gains can change both in amplitude and phase because of instrumental and also ionospheric effects. 
For a single snapshot and frequency channel, we can then write the model for the voltage output of the array as
\begin{equation}
\by[n] = \sum_q \bG_q \bs_q[n] + \bn[n],
\end{equation}
where $n=1,\dots, N$ is the sample index, $\by[n]$ is a $P \times 1$ vector obtained by stacking the output of each receiver,  $q=1,\dots,Q$ is the cluster index,  $\bG_q=\diag(\bg_q)$ is a $P \times P$ diagonal matrix with $\bg_q$ as its diagonal representing the gains (of all receivers) for the $q$th cluster, $\bs_q[n]$ is a $P\times 1$ vector representing the sampled signal from sources in $q$th cluster and $\bn[n]$ is the total receiver noise contribution to the output, coming from the system and its electronics. We assume that the system noise and the signals from each direction are independent Gaussian random variables with zero mean. The covariance model for a single snapshot and channel follows from this assumption as
\begin{equation}
\bR = \expect{\by\by^H} = \sum_q \bG_q\bSigma_q\bG_q^H + \bR_{\bn},
\end{equation}
where $\bSigma_q = \expect{\bs\bs^H}$ is the covariance of the signal from $q$th cluster (or "direction") and $\bR_{\bn} = \expect{\bn\bn^H}$ is the noise covariance matrix. An entry of the covariance matrix for two receivers is called a (complex) visibility and the vector connecting two receivers is called a baseline. There could be many visibility samples for a particular baseline pair as function of time and frequency.
Due to various model imperfections such as the existence of radio frequency interference near some receivers, it is common to remove some receiver pairs from the data \citep[e.g.][]{offringa2012morphological, 2013A&A...549A..11O, Offringamwa, 2016arXiv161004696S}. We denote by $\bM$ a symmetric masking matrix which consists of zeros and ones and is used to indicate which pairs are removed. We also remove entries where $\bR_{\bn}$ is dominant, which for uncorrelated noise are the diagonal elements, called the auto-correlations (or the zero-baseline/spacing visibilities).
A noisy estimate of the reduced covariance matrix can be directly obtained from the $P \times N$ data matrix collected during a single snapshot observation. This estimate is called a sample covariance matrix and defined as
\begin{equation}
\bRh = \bM \odot \frac{1}{N}\sum_{n=1}^N \by[n]\by[n]^H,
\end{equation}
where $\odot$ is the Hadamard or element-wise product.
Let \mbox{$k = 1,\dots,K$} be the index for each measured frequency channel and \mbox{$t=1,\dots,T$} be the index for each time snapshot, then we denote the measured sampled covariance matrix as $\bRh_{t,k}$ and the corresponding model as~$\bR_{t,k}$.
We assume that the gains remain constant during several snapshots based on the instrument and cluster. In this paper we take this time to be equal among all clusters and exactly $T$ snapshots, although this is not necessary. The gains can also be assumed constant for several adjacent frequency channels, however this is not used in this analysis and the frequency behavior of the gains is treated as a continuous and "smooth" function which we will discuss in the following sections in more details.
To summarize, our final model for a single channel and frequency is given by
\begin{equation}
\label{eq:final_model}
\bR_{t,k} = \bM_{t,k} \odot \sum_q \bG_{q,k}\bSigma_{q,t,k}\bG_{q,k}^H, 
\end{equation}
for $k = 1,\dots,K$, $t=1,\dots,T$. Note that $\bR_{\bn}$ causes a bias on the visibilities in the model and it is also a large contributor to the noise on all sampled visibilities. The contribution of $\bR_{\bn}$ as a bias is removed by $\bM$. During the remainder of this analysis, we use the vectorized form of these matrices, which are defined as
\[
\begin{array}{cc}
\br = \begin{bmatrix}
\vect(\bR_{1,1}) \\
\vect(\bR_{2,1}) \\
\vdots\\
\vect(\bR_{T,K}) 
\end{bmatrix}, &   \brh = \begin{bmatrix}
\vect(\bRh_{1,1}) \\
\vect(\bRh_{2,1}) \\
\vdots\\
\vect(\bRh_{T,K}) 
\end{bmatrix}.
\end{array}
\]
Based on the model in Eqn.~\eqref{eq:final_model} we can now formulate the calibration problem in the next section.

\section{Direction Dependent Gain Calibration}\label{sec:calib}

Calibration is the process of estimating the complex gains \mbox{$\bg_{q,k} = \vectdiag(\bG_{q,k})$} based on the measured data $\bRh_{t,k}$ and the known sky model $\bSigma_{q,t,k}$. As stated the sky model can be incomplete. The calibration procedure we address here is a least squares optimization problem of the form
\begin{equation}
\label{eq:Matrix_ls}
\bthetah = \arg \min_{\btheta} f(\btheta),
\end{equation}
where \mbox{$f(\btheta) = \|\brh - \br(\btheta)\|_2^2$} and
\[
\btheta = [\bg_{1,1}^T,\bg_{1,1}^H,\bg_{2,1}^T,\bg_{2,1}^H, \dots,  \bg_{Q,K}^H \bg_{Q,K}^H]^T
\]
is an augmented vector collating all the unknown parameters. An augmented vector is a vector where a complex variable and its conjugate are stacked together as two independent variables and it is used extensively in signal processing literature \citep[see e.g.][ and references therein]{schreier2010}. We can also formulate this problem as a weighted least square optimization similar to \citet{wijnholds2009,mouri2014}. For this analysis we assume the noise behavior of the visibility samples to be identical, which makes such a weighted formulation unnecessary. It is trivial to adapt the results to a weighted least square analysis.
The problem at hand is non-linear and non-convex and several iterative approaches have been proposed to solve it \citep[e.g.][]{aj02calibration,wijnholds2009,kazemi01102013,smirnov2015radio,mouri2014}. In this paper we do no focus on solving the gains, but to discuss possible physical constraints on the properties of the solution, $\bthetah$, once obtained. However, gains are not the only product of the calibration problem. The final residual \mbox{$\be=\brh - \br(\bthetah)$} is the target of interest for many scientific research such as the study of extremely faint cosmological 21-cm signals. It is therefore crucial to have a good understanding of the properties of $\be$. In this analysis we are interested in possible bias in the form of suppression and the associated baseline-delay (power) spectrum.

\subsection{Residuals after Least-Squares Gain Calibration}

By taking a closer look at the model, we can show that it satisfies the following property
\[
\br(\btheta) = \frac{1}{2}\bJ(\btheta)\btheta,
\]
where $\bJ(\btheta)$ is the Jacobian matrix, defined by
\[
\bJ(\btheta) = \frac{\partial \br(\btheta)}{\partial \btheta^T}.
\]
We call this property "semi-linearity" because it leads to results that appear very similar to the linear least squares problems, as we show in this section. The Jacobian itself is linear in $\btheta$ and \mbox{$\bJ(\btheta_1)\btheta_2 = \bJ(\btheta_2)\btheta_1$}. At the solution, $\bthetah$, the gradient of the cost function must vanish, and as a result we have
\begin{equation}
\label{eq:gradient_zero}
\left .\frac{\partial f(\btheta)}{\partial \btheta}\right |_{\btheta=\bthetah} = \bJ(\bthetah)^H\be(\bthetah) = \zeros.
\end{equation}
Using the property above we obtain 
\[
\bJ(\bthetah)^H\bJ(\bthetah) \bthetah = 2 \bJ(\bthetah)^H\brh,
\]
which except for the factor two on the right-hand side, is a standard normal equation. Inserting this result back into the residual, we obtain
\begin{equation}
\label{eq:residual_proj}
\be(\bthetah) = [\bI - \bJ(\bthetah)\bJ(\bthetah)^\dagger]\brh \equiv \bP(\bthetah)^\bot \brh,
\end{equation}
where $^\dagger$ is the Moore-Penrose pseudo inverse. For any matrix the product $\bP = \bA\bA^\dagger$ is an orthogonal projection into the column space of the $\bA$ and $\bP^\bot \equiv \bI-\bP$ is the projection into its null space (for more properties of general inverse matrices see e.g.\ \citet{rohde1965generalized}). For a linear problem this projection matrix is constant while in our semi-linear case it is a function of the solutions at stationary points of $f(\btheta)$.
Now we can use Eqn.~\eqref{eq:residual_proj} to study what happens to unmodeled signals in the data. To this end we extend the model for $\brh$ with several components that are the signal of interest (e.g.\ the 21-cm signal) plus another unmodeled part of the data (e.g\ diffuse foregrounds or unmodeled compact sources). Let the true gain solutions be denoted by $\bthetat$, then we have
\[
\brh = \br(\bthetat) + \br_f(\bthetat) + \br_{\text{21}}(\bthetat) + \bepsilon,
\]
where $\br_f(\bthetat)$ are any unmodeled foreground signals, $\br_{\text{21}}(\bthetat)$ is the unmodeled 21-cm signal and $\bepsilon$ is the finite sample noise on the visibilities. Putting this model into the expression for the residuals at the solution, we obtain
\begin{equation}
\be(\bthetah) = \bP(\bthetah)^\bot \left[\br(\bthetat) + \br_f(\bthetat) + \br_{\text{21}}(\bthetat) + \bepsilon\right].
\end{equation}
This result can be interpreted as follows:
\begin{itemize}
\item[(1)] The first term $\bP(\bthetah)^\bot \br(\bthetat)$ is the calibration model leakage as the result of the non-linearity of the problem, noise and model incompleteness. We note that the calibration parameters were estimate using an incomplete model of the sky and are therefore often biased with respect to the truth. This is the main mechanism which causes power of the sky to "leak" into the residuals e.g.\ \citep{2017MNRAS.470.1849E,barry2016calibration}.
\item[(2)] The second term $\bP(\bthetah)^\bot \br_f(\bthetat)$ is the remaining unmodeled foreground signals. There are nonparametric methods that can remove this part in post processing \citep{mertens2018statistical}. However, the introduced bias on the gain solutions due to this term cannot be corrected for in post processing.
\item[(3)] The third term $\bP(\bthetah)^\bot \br_{\text{21}}(\bthetat)$ is what remains of the desired 21-cm signal after the calibration, which makes the suppressed part of the signal exactly $\bP(\bthetah) \br_{\text{21}}(\bthetat)$. 
	%In other words if we consider the residuals as an estimate of the $\br_{\text{21}}$, the error on this estimate is 
	%\[
	%\text{Error}_{eor} = \bP(\bthetah)^\bot\left[\br(\bthetat) + \br_f(\bthetat) + %\bepsilon\right]-\bP(\bthetah)\br_{\text{21}}
	%\]
\end{itemize}
We would like to minimize the first two terms while keeping the third term intact as much as possible. We see that this leads to conflicting objectives. If we model and include sources in $\br_f$ (by adding them to $\br$), we reduce the model incompleteness and the leakage from the first two terms. However, doing so could increases the degree of freedom in the model which allows for removing even a larger part of the desired signal. In other words, it could increases the rank of the projection matrix, $\bP(\bthetah)$, and remove more power from $\br_{\text{21}}$. If based on physical arguments, the solution space can be restricted further, then we can also reduce the signal suppression.
In fact, there exists a strong physical justification for assuming that the gains are smooth functions of the frequency \citep{vandertol2007}, which we will discuss next. 

\subsection{Enforcing Smoothness of the Gains}\label{ssec:smooth}

In the following we assume two scenarios: in both cases we assume that the gain solutions follow a smooth functional form (e.g.\ polynomial), where in the first case the higher order (i.e.\ those describing gain fluctuations on smaller frequency scale) terms are not regularized, whereas in the second case they are regularized.  

\subsubsection{Smoothness of Gain via Basis-Function Constraints}

We consider enforcing smoothness of the gains by means of choosing a set of parametric smoothing basis functions.  Let the vector $\blambda$ be a $K \times 1$ vector with the central frequency of each channel as its elements. Then we denote by $\bPhi(\blambda)$ a $K \times M$ unitary basis for the set of sampled smooth functions. For example, for a polynomial of order $m$, we have $M = m + 1$ and $\bPhi(\blambda)$ can be obtained from an economical QR or SVD on the following Vandermonde matrix
\[
\bV =  \begin{bmatrix}
1 & \lambda_1 & \lambda_1^2 & \dots & \lambda_1^m \\
& & \vdots & & \\
1 & \lambda_K & \lambda_K^2 & \dots & \lambda_K^m \\
\end{bmatrix},
\]
where we assume $M < K$. This can be done with {\sl any} set of smooth basis functions, e.g.\ Bernstein polynomials as in \citet{Yatawatta:2016fw}. We will drop the dependency on $\blambda$ from the notation as it is assumed not to change during a single calibration run of $T$ snapshots. Using $\bPhi$ we can model the gains as
\begin{equation}
\label{eq:smooth_model}
\btheta = (\bPhi \otimes \bI_{2PQ})\balpha,
\end{equation}
where the entries of $\balpha$ are the coefficients for the smooth functions, $\bI_{2PQ}$ is an identity matrix of size $2PQ$ and $\otimes$ is the Kronecker product. Note that the only change in the analysis of the previous section is the definition of the Jacobian which then becomes 
\[
\bJ_s(\balpha) = \bJ(\btheta) (\bPhi \otimes \bI_{2PQ}).
\]
The semi-linearity is retained by this change of variable, while the column dimension of $\bJ$ is reduce by a factor of $K/M$ (e.g. for LOFAR this factor can be more than a hundred for calibration of hundreds of channels with a low order polynomial). If accurate, this model should reduce the 21-cm signal suppression substantially, while not increasing the leakage from the sky model into the residuals. The foreground leakage (i.e. the unmodeled nuisance part of the data which is not of interest to the analysis of the desired signal) can increase, but as before we assume this to be solved in a post processing step.

\subsubsection{Regularization of High-order Gain Model Terms}
\label{ssec:regularization}

Direction dependent calibration is typically done on time scales of several minutes (e.g.\ $\sim$\,$10$ minutes for LOFAR) for a data set of hundreds of hours. This means the variation in gains for different clusters, especially the ones further from the center of the main beam, can be large. Choosing a fixed basis function might not be enough. In this scenario a possible solution is to allow wider set of basis functions, or equivalently enlarge $M$ and allow for more flexibility in $\bPhi$, while penalizing the newly added dimensions with a regularization term. We will investigate this here. Let \mbox{$M = M_1 + M_2$} such that we can split the enlarged $\bPhi$ as 
\[
\bPhi =[\bPhi_1,~\bPhi_2],
\]
where $\bPhi_2$ is the extra added freedom. In this case we have \mbox{$\btheta = (\bPhi_1 \otimes \bI_{2PQ})\balpha_1 + (\bPhi_2 \otimes \bI_{2PQ})\balpha_2$}, which is a smooth function with an additional "less" smooth variation around it. By limiting the magnitude of $\balpha_2$, we control this additional term. We can formulate this approach as a new optimization problem
\[
\begin{array}{rl}
\bthetah &= \arg\min_{\btheta} f(\btheta) + \sum_q \gamma_q \btheta_q^H(\bPhi_2\bPhi_2^H \otimes \bI_{2P})\btheta_q \\
&\text{s.t. } \bP_{\bPhi}^\bot \btheta = \zeros,
\end{array}
\]
where \mbox{$\bP_{\bPhi} = \bPhi\bPhi^H \otimes \bI_{2PQ}$}, $\gamma_q$ is a regularization parameter and $\btheta_q$ is the subset of $\btheta$ corresponding to $q$th cluster. This model still enforces a smooth solution of a polynomial of (maximally) order $M-1$, but if the regularization is large the solution reduces back to a polynomial of order $M_1-1$, as in the case discussed in previous section. Standard regularization (e.g.\ Tikhonov) around a smooth function would be equivalent to letting $M_2 = K - M_1$, which could be a large number (depending on the number of frequency channels). 
The regularization parameters can be chosen based on the statistics of the gains, if they are known, or by updating them during optimization. Here we assume that the regularization parameters, while (possibly) different for each cluster, are constant values during a single calibration run of $T$ time samples. The constraint \mbox{$\bP_{\bPhi}^\bot \btheta = \zeros$} is a different way of stating Eqn.~\eqref{eq:smooth_model} in the optimization setting, forcing the solution to be a low-order polynomial, while \mbox{$\btheta_q^H(\bPhi_2\bPhi_2^H \otimes \bI_{2P})\btheta_q$} suppresses the contribution of $\bPhi_2$ to the solution depending on the level of regularization (i.e.\ values of $\gamma_q$).
Using the semi-linearity again, we can show that in this case the residuals are given by \mbox{$\be(\bthetah) = [\bI - \bZ(\bthetah)]\brh$} where 
\[
\bZ(\btheta) =  \bJ(\btheta)\bP_{\bPhi} \left[\bJ(\btheta)^H\bJ(\btheta) + 2\bGamma(\bPhi_2\bPhi_2^H \otimes \bI_{2PQ})\right]^\dagger\bP_{\bPhi} \bJ(\btheta)^H,
\]
and $\bGamma$ is a diagonal matrix with $QK$ blocks of the form $\bgamma_q \bI_{2P}$. Note that, due to regularization, the matrix $\bZ$ is not a projection. However, the expression for the residual still allows us to calculate the suppression exactly for a given solution $\bthetah$, the basis functions in $\bPhi$ and a set of regularization parameters $\gamma_q$ in a similar way to the previous section. The interpretation of leakage terms remains valid.
Given the large number of calibration runs that must be performed for long observations, it is important to know if we can relax the stopping criteria for applied algorithms. Here we assume to have access to an algorithm that solves smooth solution by using a single basis set equivalent to the scenario in previous section (i.e. $M=M_1$). We also assume that the chosen algorithm does not enforce the constraint \mbox{$\bP_{\bPhi}^\bot \btheta = \zeros$} at each iteration and it is terminated before full convergence. By translating this constrained problem to a regularized problem, we can approximate its behavior. We allow $\bPhi_2$ to "fill" the entire parameter space or equivalently let $M_2 = K - M_1$. For this regularized problem we have \mbox{$\bP_{\bPhi} = \bI$} and \mbox{$\bPhi_2\bPhi_2^H = \bI-\bPhi_1\bPhi_1^H$}. The entire regularization is then specified by $\bPhi_1$, which must be the case (because the original problem assumes $M_2$ to be zero). This allows for more fluctuations in the solutions, but  because of the regularization, the magnitude of these fluctuations is limited. Hence, by increasing the regularization parameters, we simulate the effect of increasing the number of iterations.  In the limit, the solution will be forced on the subspace of $\bPhi_1$ which is also what happens at convergence of the original problem.
\subsection{Effect of Baseline Cut on the Residuals}
One strategy to avoid a possible suppression of the desired 21-cm signal is to exclude the baselines on which we want to preserve the signal from the calibration data. \citet{patil2017upper} follow such a strategy by removing all the baselines below 250 $\lambda$ from the calibration data. Although this indeed eliminates suppression on shorter baselines, it also causes an increased power, denoted by ``excess noise", on the residuals for the excluded baselines. In this section we derive the expression for the residuals on the excluded baseline and in the next section we show, by simulations, how this strategy affects the power spectrum.
While it is possible to include the baseline cut in the mask matrix, $\bM$, the expression for the residual given in previous sections would then only be valid for the visibilities which are included during the calibration. In fact we are mainly interested in the residuals on the removed baselines and hence we reformulate this problem in an slightly different way.

\noindent
The calibration problem with a baseline cut can be formulated as 
\[
\begin{array}{rl}
\bthetah &= \arg\min_{\btheta} \|\bP_{sb}^\bot [\brh - \br(\btheta)]\|_2^2 + \sum_q \gamma_q \btheta_q^H(\bPhi_2\bPhi_2^H \otimes \bI_{2P})\btheta_q \\
&\text{s.t. } \bP_{\bPhi}^\bot \btheta = \zeros,
\end{array}
\]
where $\bP^\bot_{sb}$ is a projection matrix that removes the short baselines from the data. This optimization problem is a slightly modified version of the problem discussed in Sec.~\ref{ssec:regularization}. Using the semi-linearity, the residuals on the removed baselines are found to be
\begin{equation}
\be(\bthetah) = \bP_{sb}[\bI - \bZ_{sb}(\bthetah)]\brh,
\end{equation}
where
\[
\bZ_{sb} = \bJ\bP_{\bPhi} \left[\bJ^H\bP_{sb}^\bot\bJ + 2\bGamma(\bPhi_2\bPhi_2^H \otimes \bI_{2PQ})\right]^\dagger\bP_{\bPhi} \bJ^H\bP_{sb}^\bot,
\]
and we have dropped dependency on $\btheta$ here for better readability. The main difference between the expression for $\bZ_{sb}$ and $\bZ$ from the previous sections is the term $\bJ^H\bP_{sb}^\bot\bJ$. Ignoring the regularization term for the moment, we see that this matrix indicates the level with which the gains can be calibrated independently of the removed baselines. If this term is singular, it indicates that the (unconstrained) problem is not identifiable and if it is badly conditioned, it will result in signal amplification. In the next section we see that this is exactly what happens, especially if the sky-model errors are large.
In order to gain more insight into the suppression behavior of gain calibration, we need to test these results with a sufficiently realistic model. That is the objective of the next section.

\section{Simulations of 21-cm Signal Suppression} \label{sec:sim}

Given a realistic simulation of an instrument, we can directly calculate the suppression of any 21-cm signal during the calibration process, in the context of a calibration model that consists of multiple source clusters, an unmodeled component of the sky, and the desired 21-cm signal plus noise. In this section we use LOFAR \citep{lofar_ref} and the LOFAR EoR KSP calibration approach \citep[see][for details]{patil2017upper} as an example for testing the difference between different calibration approaches.
We study the suppression using a Gaussian random field for the 21-cm signal which is maximally white (spatially and in frequency) in order to see the effect equally well on all scales of interest. Given that the signal suppression is scale dependent but relative, the exact choice of the 21-cm signal is of secondary nature since it drops out when calculating the ratio between the input and output signal and only if chosen too strong might it affect the gain solutions.

\subsection{Simulation Setup}

For the following simulations we have $P=62$ receivers (LOFAR stations), a model of the North Celestial Pole, which is one of the fields used for 21-cm EoR signal detection with LOFAR-HBA. The model is currently used for calibration of LOFAR telescope \citep{patil2017upper}. We have simplified the model slightly by replacing some compact Gaussian sources with point sources of equal magnitude, which is not a relevant change. We use $K=53$ sub-bands with a bandwidth of 195.3 kHz which gives us approximately a total bandwidth of 10 MHz centered symmetrically around 150 MHz. The 28000+ source components are split into $Q=122$ clusters spread around the sky, but predominantly inside the primary beam of LOFAR-HBA. For simplicity, although this is motivated by the slow changes in gains in LOFAR, we assume the gains to remain constant (or coherent) during a period of 10 minutes and the data to have a time resolution of 1 second which gives $T=600$. Sampling at the Nyquist rate, we have $N= \lceil 2 \times 195.3 \times 10^3 \rceil$ electric field measurements for each snapshot of 1 second and for each receiver. An additional noise term is added to account for the receiver noise and sky temperature using $\bR_{\bn}$. We simulate the 21-cm signal only on baselines with a length less than 250 $\lambda$. These baselines are also used to make the final baseline-delay power spectra. On the longer baselines the signal is too weak to be detected by LOFAR.
The algorithm used is a Newton-based algorithm with a stopping criteria of \mbox{$\|{\partial f(\btheta)}/{\partial \btheta}\| < 10^{-8}$} which is small enough for $\bthetah$ to be a critical point of the cost function and to ensure that the relations used for suppression are valid. The technical details of the algorithm are beyond the scope of this paper and will be reported separately.
For the smooth basis functions we use a polynomial set. The maximum allowed freedom for these polynomials is chosen to be 3 which means that $M=4$. We chose the large scale fluctuation of the gains for 10 MHz to be dominated by a first order polynomial, which makes $M_1=2$.

\subsection{Suppression Test using the Complete Sky Model}

\begin{figure}
	\begin{center}
		\mbox{\epsfig{figure=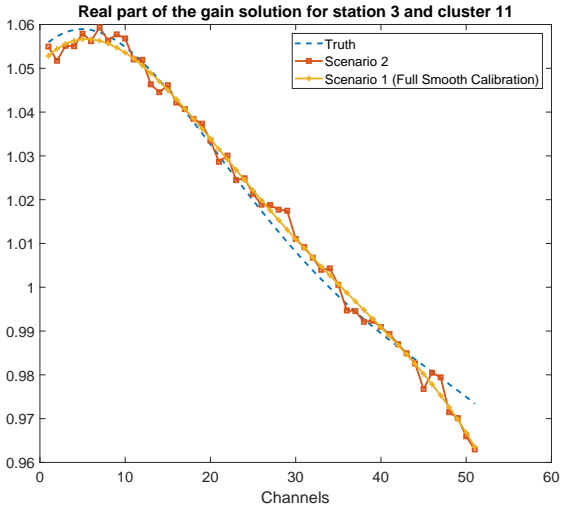,width=.49\textwidth}}
	\end{center}
	\caption{Real part of the gain solutions as a function frequency for a single station and cluster.}
	\label{fig:solutions}
\end{figure}

\begin{figure}
	\subfloat[Partially enforced smoothness.]{
		\includegraphics[width=.49\textwidth]{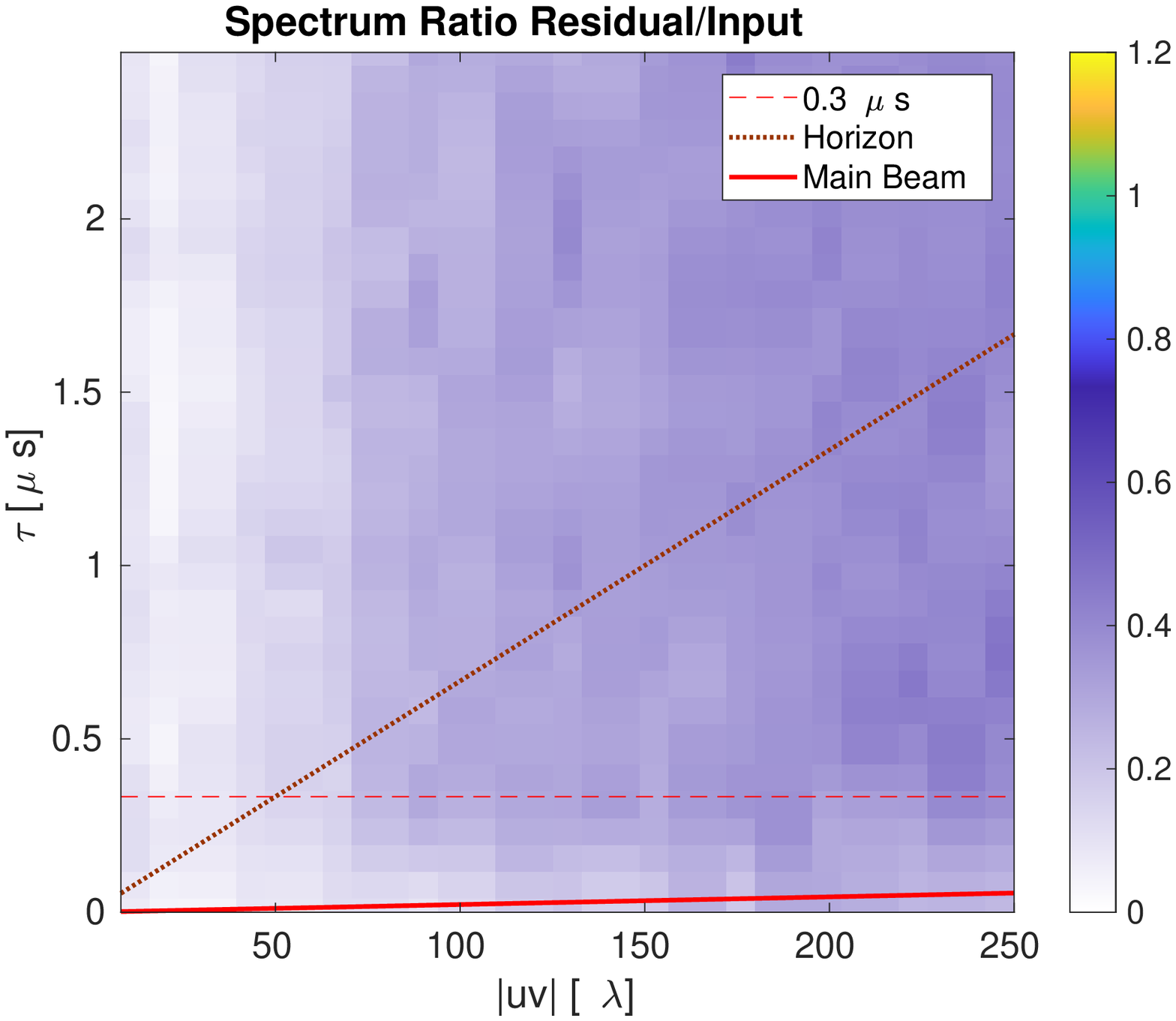}
	}
	\hfill
	\subfloat[Fully enforced smoothness.]{%
		\includegraphics[width=.49\textwidth]{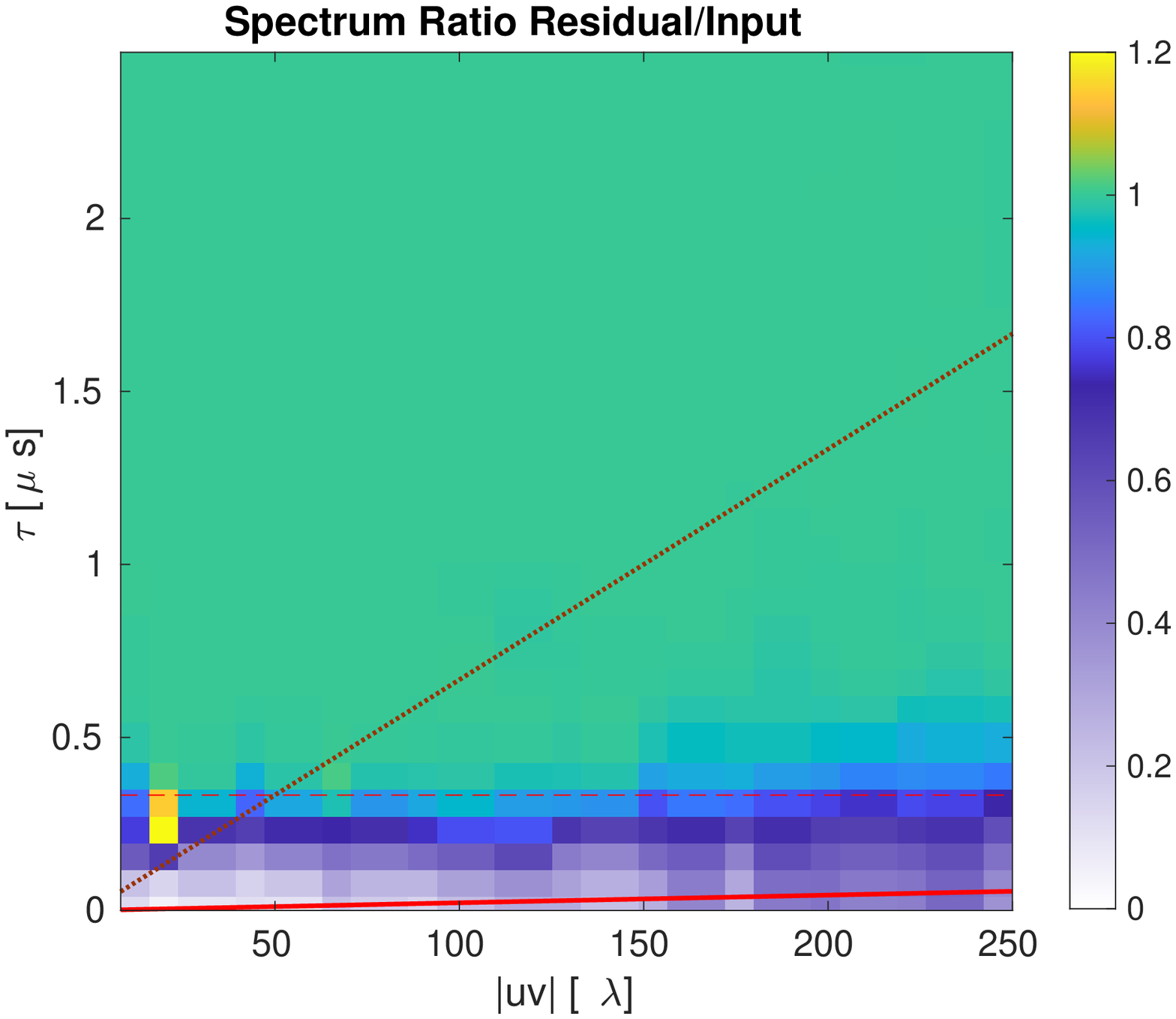}
	}
	\caption{Ratio of power spectra for the output and the input 21-cm signals. The top panel shows their ratio when a complete sky model is used in the calibration, but allowing for some residual fluctuations of the gains around a smooth model, mimicking non-convergence or regularization. The bottom panel forces the solutions exactly on a smooth model of low order.}
	\label{fig:ratios_complete}
\end{figure}

We test the 21-cm signal suppression in two scenarios. In one scenario we do a full smooth estimation (Scenario\,1) and in the other scenario (Scenario\,2) we study the effect of enforcing only $\bPhi_1$, but stopping before convergence as discussed in Sec.~\ref{ssec:smooth}. This means that in Scenario\,1, $M_1 = 2$, $M_2 =2$ while in Scenario\,2 we have $M_1 = 2$ and $M_2 = K - M_1$. We choose a much larger regularization term for the latter in order to emulate premature termination after several iterations. 
The sky-model is used for both generation of the data as well as calibration and hence except for the 21-cm signal and noise there are no addition signals, i.e. $\br_f=\zeros$. The matrix $\bV$ is constructed and then, via an economical QR decomposition, $\bPhi$ is obtained. The gains are generated as sum of two components for each part of $\bPhi$. The expected value of the gains is generated with $\bPhi_1$ and a proper complex normal distributed random variables with a variance $\sigma^2 = 0.07$ is generated for $\bPhi_2$. The regularization terms are then chosen uniformly with $\gamma_q = 1/\sigma^2$ for the full smooth calibration in Scenario\,1. For the second simulation, Scenario\,2, we use $\gamma_q = 100/\sigma^2$ to account for several iterations that forces the solutions closer to $\bPhi_1$, but not exactly on the subspace spanned by $\bPhi_1$. This is similar to Scenario\,1, but mimics terminating optimization before convergence, hence leaving extra gain fluctuations from $\bPhi_2$ in the solution.
Fig.~\ref{fig:solutions} illustrates an example of the difference between the solutions for the two scenarios. As expected the solution for for Scenario\,1 is much smoother and as a result has a much smaller degree of freedom (i.e. less dimensions are projected out). This is confirmed by the gain-solution results illustrated in Fig.~\ref{fig:ratios_complete}. The latter figure clearly shows that if smoothness is not exactly enforced, a large portion of the 21-cm signal power-spectrum is heavily suppressed (up to an order of magnitude). It is important to note that if the suppression is expressed in the terms of the 2-norm of the signal $\|\bZ\br_{\text{21}}\|_2/\|\br_{\text{21}}\|_2$ only a suppression of 17\% is estimated for Scenario\,2. This indicates that calculating the power spectrum, in the region of interest, gives a much more accurate picture of suppression. Suppression is stronger on the shorter baselines and smaller delays. We note that this suppression on short baselines is similar to that seen by \citet{2016MNRAS.463.4317P}. 
The polynomial basis function used is of order 3, undulating on scales of $\approx 0.3\mu$s or smaller. Any scale below this can therefore be modeled by this basis function, and it is exactly where the suppression increases rapidly.

\subsection{Suppression Test with an Incomplete Sky Model}

\begin{figure}	
	\subfloat[Scenario 2 the smoothness partially achieved.]{
		\includegraphics[width=.49\textwidth]{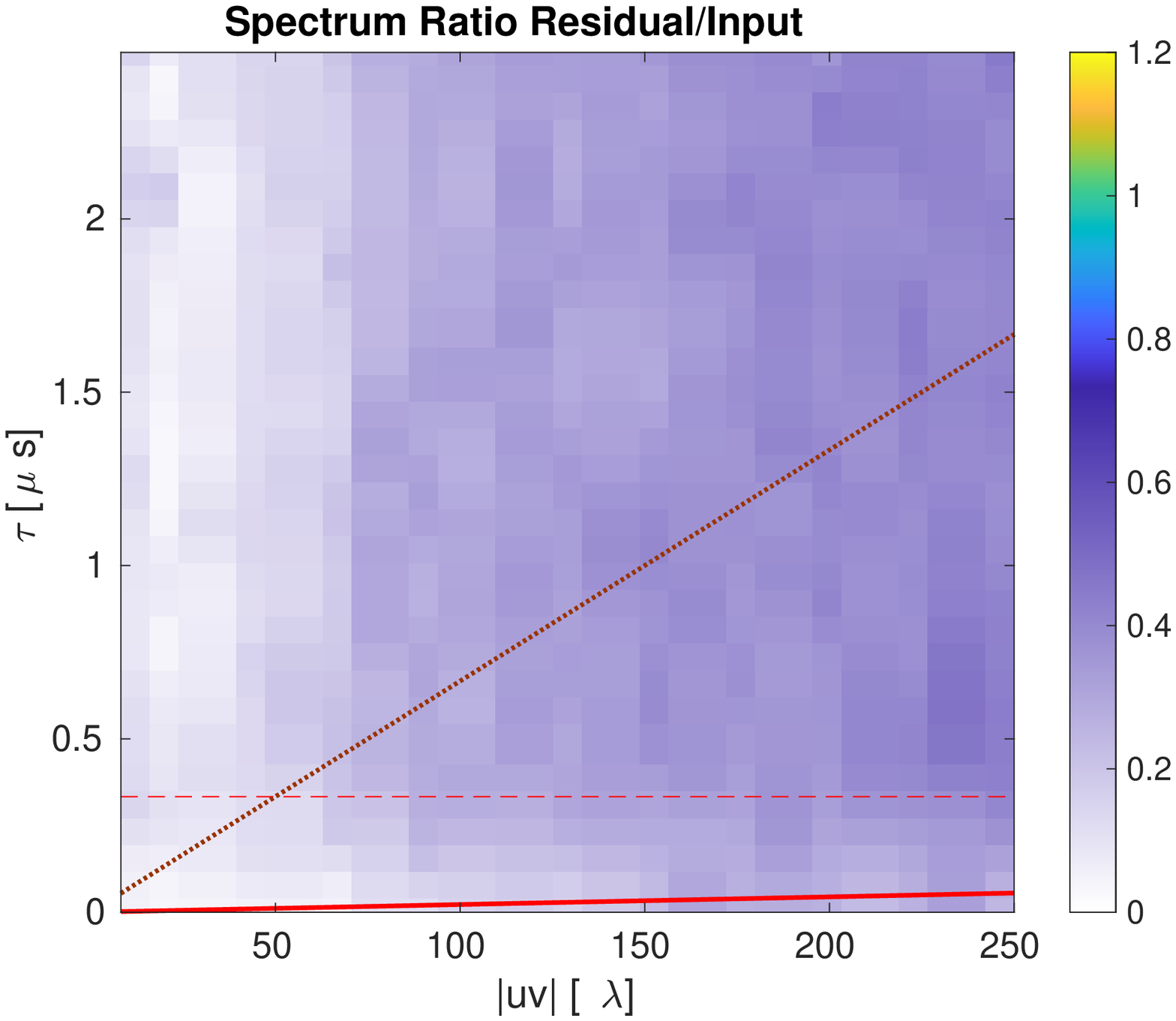}
	}
	\hfill
	\subfloat[Scenario 1 the smoothness is fully enforced]{%
		\includegraphics[width=.49\textwidth]{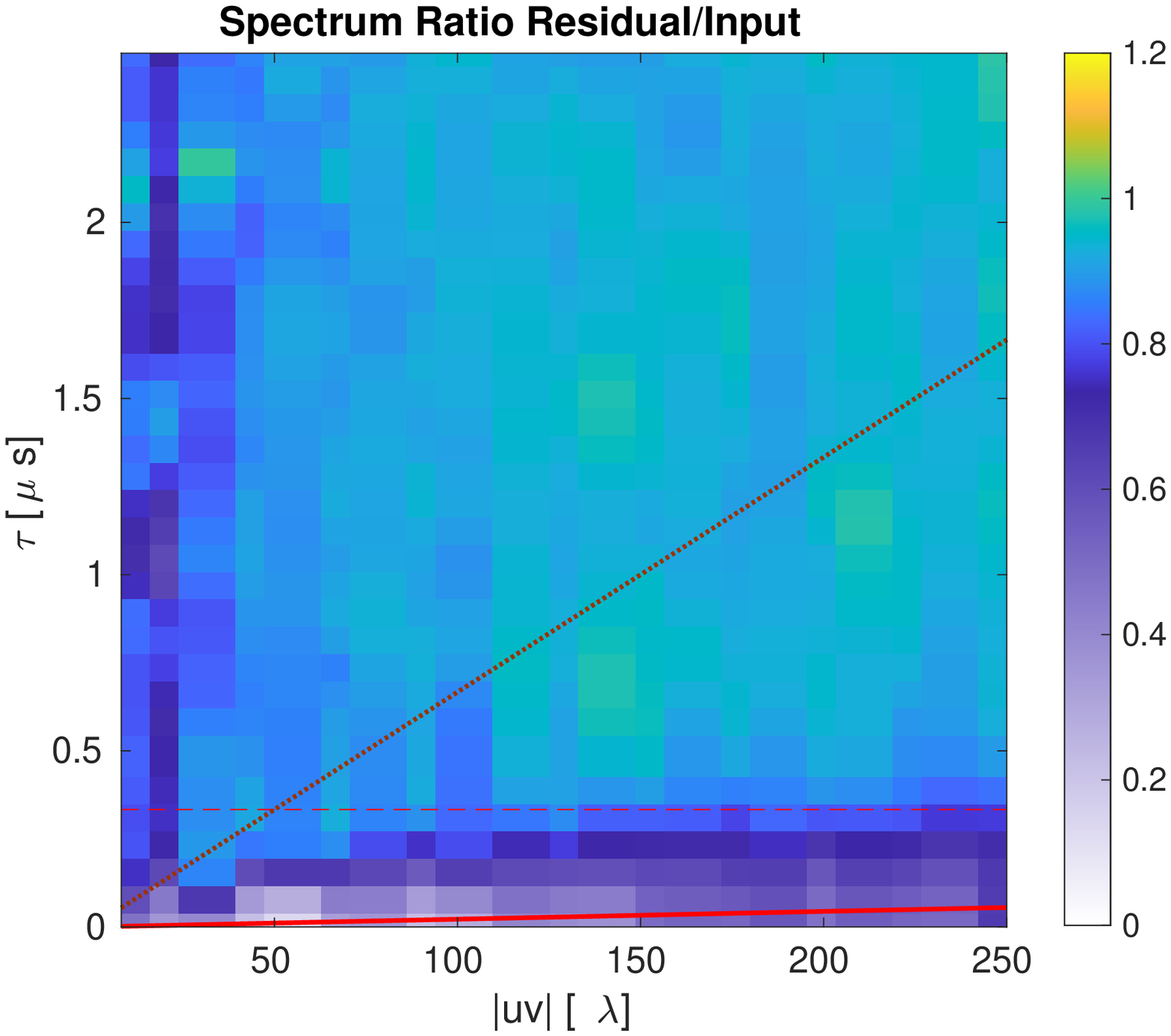}
	}
	\caption{Ratio of power spectra for the output and the input 21-cm signals. The top panel shows their ratio when a incomplete sky model is used in the calibration, but allowing for some residual fluctuations of the gains around a smooth model, mimicking non-convergence or regularization. The bottom panel forces the solutions exactly on a smooth model of low order.}
	\label{fig:ratios_incomplete}
\end{figure}

The final simulation is that where the sky-model is incomplete, which is much closer to a realistic calibration scheme.
The setup for this simulation is similar to the one in the previous section with the exception that all the sources smaller than 1~mJy are removed from the calibration model. This adds an additional foreground term and $\br_f \ne \zeros$. Again we calculate the power spectra for both scenarios and compute the suppression. 
Fig.~\ref{fig:ratios_incomplete} shows that by using an incomplete sky model the suppression is more prominent even when we fully enforce smoothness. This can be explained by using the results in Appendix~\ref{app:1} which show that the suppressed part of the 21-cm signal is proportional to the error in the model, \mbox{$\br(\bthetat) - \br(\bthetah)$}, and the foreground term $\br_f$, which are both larger in this scenario. Hence sky-incompleteness in the calibration model leads to enhanced 21-cm suppression.

Although, to limit computational effort, we have removed sources $< 1$ mJy from the model rather than adding sources below the lower flux limit to visually demonstrate the effect of model incompleteness; We note that in reality these sources are part of the LOFAR sky-model and hence the level of suppression is likely to be smaller than illustrated in Fig.~\ref{fig:ratios_incomplete} (bottom panel).

\subsection{Suppression Test with a Baseline Cut}
In this section we will repeat all four tests done in the previous section with the additional modification of a baseline cut. All the baselines below 250 $\lambda$ are removed from the data prior to the calibration and the effect of this operation on the power spectrum is studied.

\begin{figure}	
	\subfloat[Scenario 2 the smoothness partially achieved.]{
		\includegraphics[width=.49\textwidth]{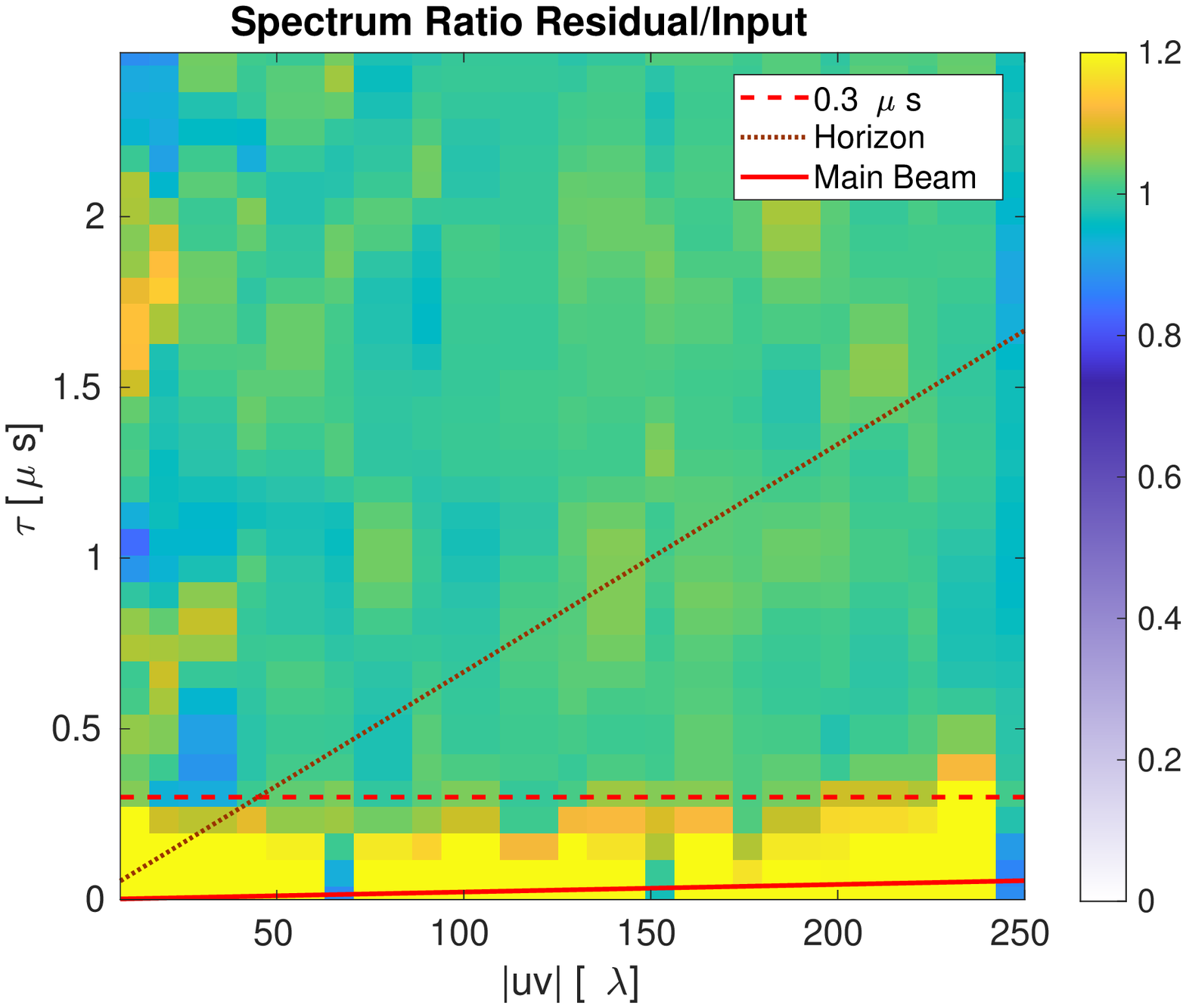}
	}
	\hfill
	\subfloat[Scenario 1 the smoothness is fully enforced]{%
		\includegraphics[width=.49\textwidth]{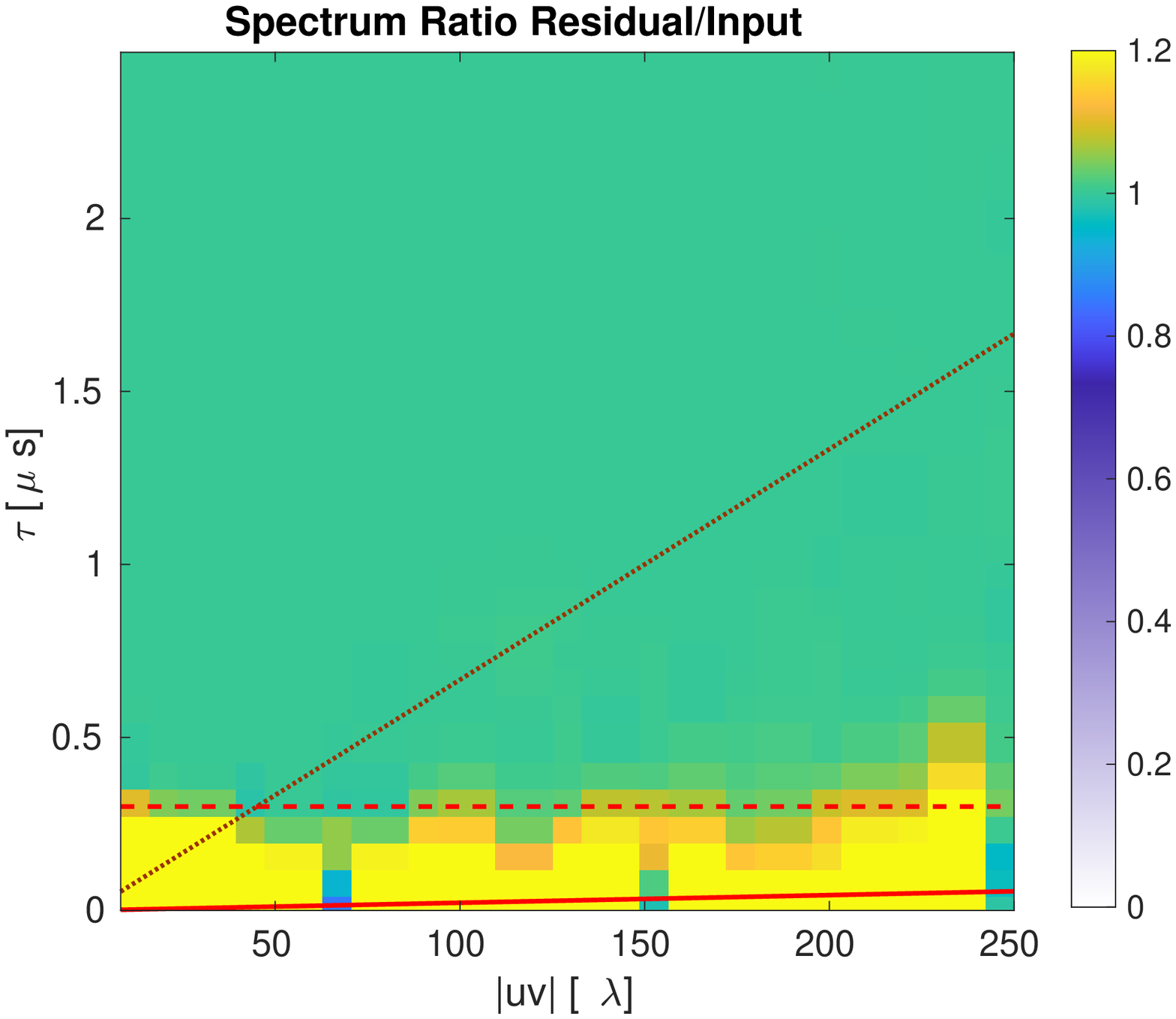}
	}
	\caption{Ratio of power spectra for the output and the input 21-cm signals. The top panel shows their ratio when a baseline cut of 250 $\lambda$ is used in the calibration, but allowing for some residual fluctuations of the gains around a smooth model, mimicking non-convergence or regularization. The bottom panel forces the solutions exactly on a smooth model of low order.}
	\label{fig:ratios_complete_cut}
\end{figure}

\begin{figure}	
	\subfloat[Scenario 2 the smoothness partially achieved.]{
		\includegraphics[width=.49\textwidth]{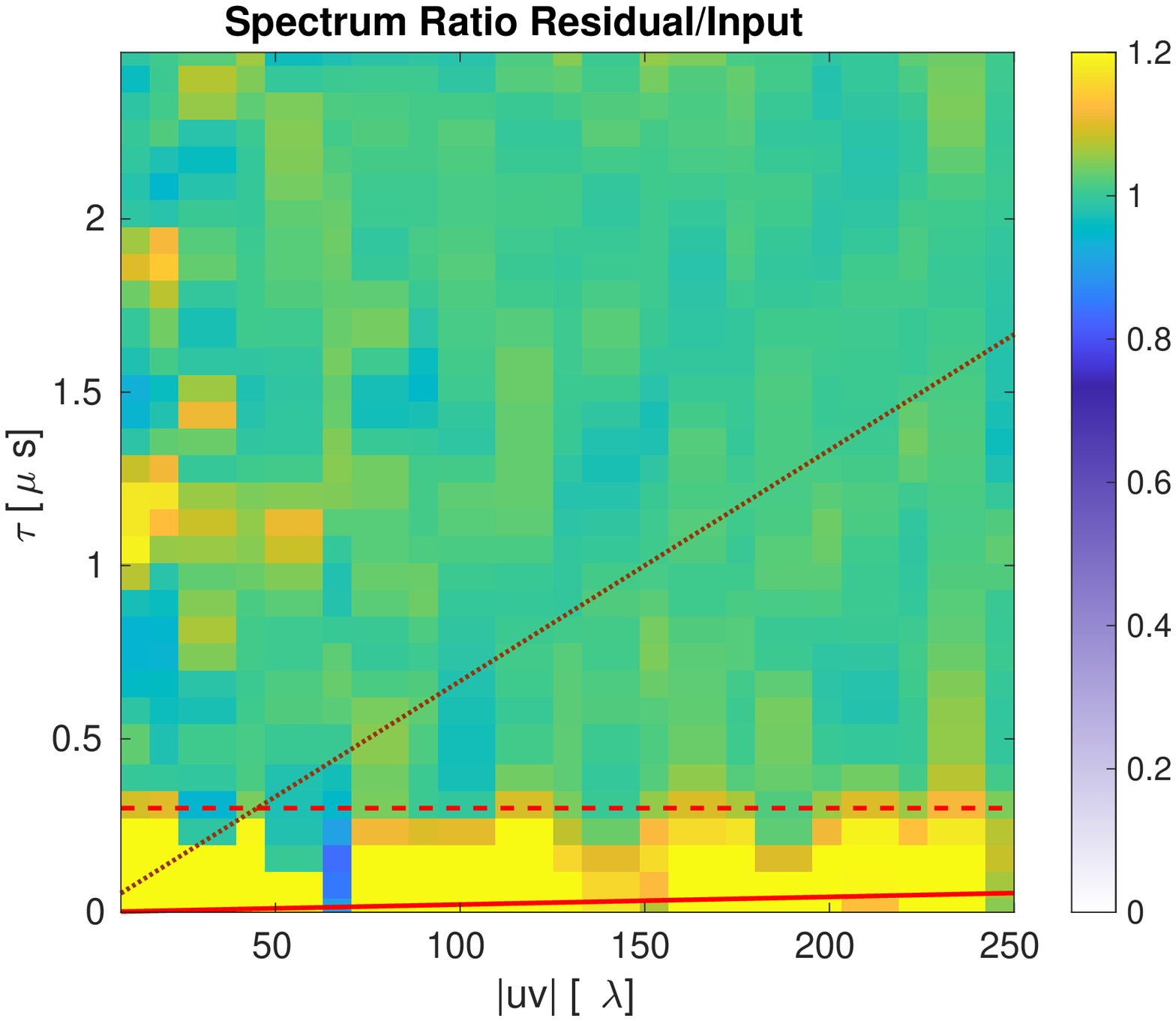}
	}
	\hfill
	\subfloat[Scenario 1 the smoothness is fully enforced]{%
		\includegraphics[width=.49\textwidth]{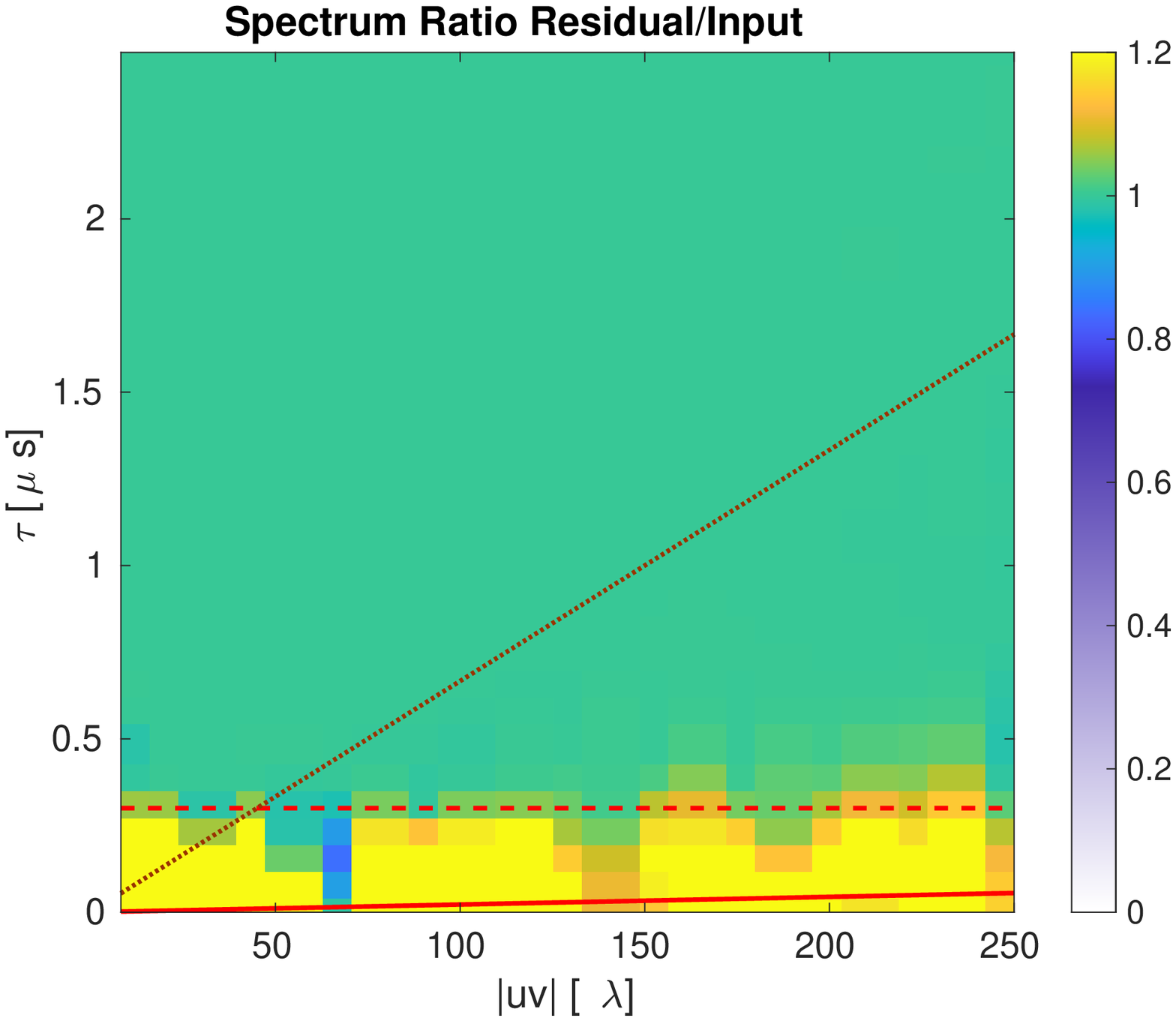}
	}
	\caption{Ratio of power spectra for the output and the input 21-cm signals. The top panel shows their ratio when a incomplete sky model a baseline cut of 250 $\lambda$ is used in the calibration, but allowing for some residual fluctuations of the gains around a smooth model, mimicking non-convergence or regularization. The bottom panel forces the solutions exactly on a smooth model of low order.}
	\label{fig:ratios_incomplete_cut}
\end{figure}

Fig.~\ref{fig:ratios_complete_cut} and Fig.~\ref{fig:ratios_incomplete_cut} show the ratio of the spectra for both complete and incomplete sky models. By comparing these results with the results from previous sections, we see that the suppression behavior of the algorithm in the case of partial smoothness is significantly improved. We note that this is the calibration strategy taken in \citet{patil2017upper}. We also see that where the signal was mainly suppressed, it is now amplified, which indicates degeneracy in the calibration model on those scales. This effect has also been seen in \citep{patil2017upper} and earlier described in \citep{2016PASA...33...19T}. These figures also illustrate that both with and without a baseline-cut it is beneficial to enforce smoothness more rigorously. Both the suppression and amplification of the signal (and with high probability also the leakage terms) is reduced which agrees with the results in \citep{2017MNRAS.470.1849E}.   

\section{Conclusions}\label{sect:concl}

In this paper we have presented the first theoretical model that allows one to quantify the suppression of the cosmological 21-cm signal, due to direction dependent gain calibration of radio-telescope receivers (e.g.\ dishes or aperture-arrays). 
By exploiting the ``semi-linearity" property of the gain calibration models, we have shown that this suppression can be quantified with high precision and, as long as the signal is weak and does not bias the gain solutions themselves strongly, the result is independent of the 21-cm signal itself. The closed form expressions for the residuals, that we present, allow one to study various leakage terms and clearly illustrate the trade-offs inherent to the calibration problem and the choices that are made (e.g.\ completeness of the sky model, number of directions to solve for, number of frequency channels to include, the level smoothness enforced on the solutions, baseline cut, etc.).
We have found the following main results:
\begin{itemize}
\item[(1)] If the sky model is complete, constraining and regularizing direction dependent gain solutions to be smooth functions of frequency reduces suppression of the 21-cm signal to nearly zero on all short baselines, above a delay that corresponds roughly to the scales on which the gains can vary. We demonstrate this for LOFAR, showing that its current direction dependent  gain calibration \citep[as described in][]{Yatawatta:2016fw} in just over a hundred directions, is a feasible approach as long as the solutions are enforced to be smooth on scales larger than several MHz, and assuming that also the true gains are smooth on those scales (which is an instrument requirement for {\sl any} successful 21-cm detection).

\item[(2)] To our knowledge, a completely new result is that an incomplete sky model does not only cause leakage of strong foreground signals in to the cosmological 21-cm signal as show by \citet[e.g.][]{barry2016calibration, 2016MNRAS.463.4317P}, it also degrades the quality of the 21-cm residuals {\sl and} increases signal suppression even if smoothness is fully enforced. This increases 21-cm signal suppression even if the foregrounds and the 21-cm signals are fully uncorrelated.

\item[(3)] Finally we find that introducing a base line cut, where one calibrates the signal on shorter baseline using only the longer baselines, removes signal suppression, but conversely causes a enhanced signal when the sky model is incomplete. Enforcing spectral smoothness, however, as also shown by \citet{2017MNRAS.470.1849E}, removes this signal enhancement. 
\end{itemize}

\noindent
Although we conclude that forcing the gain solutions on smooth functions is an absolute requirement for 21-cm signal detection experiments \citep[see  discussions on these requirements in][]{Yatawatta:2016fw, barry2016calibration, 2016PASA...33...19T}, as is having a sky model that is as complete as possible, in practice  this is computationally often very expensive. Performing for example 30--40 iterations of a full Newton-based calibration or increasing the sky model by an order of magnitude in number of components is not feasible when thousands of hours of data (petabytes) have to be processed. Yet both are crucial to avoid significant suppressing of the desired 21-cm signal. Adding a baseline cut mitigates suppression at the cost of enhancing power due to calibrating on an incomplete sky model. Finding the right balance between enforcing sufficient smoothness and reaching full convergence and having a sufficiently complete sky model, as attempted in the calibration approach of \citet{Yatawatta:2016fw}, is therefore essential for detecting and properly quantifying the 21-cm signal. 
We furthermore conclude that our results for LOFAR indicate that the low-frequency part of the Square Kilometre Array (i.e.\ SKA-low) -- being very similar to LOFAR, having a much improved sensitivity and instantaneous uv-coverage -- should therefore also be calibratable in at least $\sim$100 directions for each station, if the direction dependent receiver (i.e.\ station) instrument and model gains are smooth on scales of several MHz and if the sky model is complete to the mJy level, much in line with the conclusions of \citet{2016PASA...33...19T}.

\section*{Acknowledgements}

AMS and LVEK acknowledge support from a SKA-NL Roadmap grant from the Dutch ministry of OCW. We also thank Sarod Yatawatta for providing us with the LOFAR NCP sky model, and for very useful discussions and feedback. We also thank Andr\'e Offringa, Florent Mertens, Bharat Gehlot and Maaijke Mevius for useful discussions on about gain calibration and modeling.

\bibliographystyle{mnras}
\bibliography{biblio3}

\appendix
\section{Incompleteness and suppression}
\label{app:1}
The semi-linearity allows us to write the relation between the residuals at any two points $\btheta_1$ and $\btheta_2$ as
\[
\be(\btheta_2) = \be(\btheta_1) - \br(\bDelta\btheta) - \bJ(\btheta_1)\bDelta\btheta
\]
where \mbox{$\bDelta\btheta = \btheta_2-\btheta_1$}.
Two important points are the true gains, $\bthetat$ and the solution to the optimization problem $\bthetah$.
Using these two points in the relation above, we can express the removed part of the 21-cm signal as a function of other contributors to the visibilities. For the residual at the solution we have
\[
\be(\bthetah) = \be(\bthetat) - \br(\bDelta\btheta) - \bJ(\bthetat)\bDelta\btheta.
\]
While this expression shows how the residual is related to the error in the solutions, \mbox{$\bthetah-\bthetat$}, it does not show how  exactly the error in the visibility affects these solutions. We know from Eqn.\,\eqref{eq:gradient_zero} that  at the solution \mbox{$\bP(\bthetah)\be(\bthetah)=\zeros$}. By multiplying both sides of the expression above by $\bP(\bthetah)$ and writing out the expression for $\be(\bthetat)$ we have
\begin{equation*}
\bP(\bthetah)\left[\br(\bthetat) - \br(\bthetah) -\br_{\text{21}}(\bthetat)-\br_f(\bthetat)-\bepsilon\right] = \zeros
\end{equation*}
or
\begin{equation}
\bP(\bthetah)\br_{\text{21}}(\bthetat) =  \bP(\bthetah)\left[\br(\bthetat) - \br(\bthetah) -\br_f(\bthetat) - \bepsilon\right].
\end{equation}
This shows that the suppressed part of the 21-cm signal is proportional to the error in the model, \mbox{$\br(\bthetat) - \br(\bthetah)$}, which is the result of noisy/biased solutions and foregrounds. This also shows that when $\br_f$ is large, it contributes two times to the signal degradation. The first contribution is direct and the second is by increasing the bias in $\bthetah$ and hence increasing \mbox{$\br(\bthetat) - \br(\bthetah)$}.

% Don't change these lines
\bsp	% typesetting comment
\label{lastpage}
\end{document}

%% file: notations.tex
\newcommand{\vect}[0]{\text{vect}}

\newcommand{\vectdiag}[0]{\text{vectdiag}}

\newcommand{\diag}[0]{\text{diag}}

\newcommand{\bGamma}[0]{\boldsymbol{\Gamma}}
\newcommand{\bDelta}[0]{\boldsymbol{\Delta}}

\newcommand{\bSigma}[0]{\boldsymbol{\Sigma}}
\newcommand{\bPhi}[0]{\boldsymbol{\Phi}}

\newcommand{\bA}[0]{\mathbf{A}}

\newcommand{\be}[0]{\mathbf{e}}

\newcommand{\bg}[0]{\mathbf{g}}
\newcommand{\bG}[0]{\mathbf{G}}

\newcommand{\bI}[0]{\mathbf{I}}

\newcommand{\bJ}[0]{\mathbf{J}}

\newcommand{\bM}[0]{\mathbf{M}}
\newcommand{\bn}[0]{\mathbf{n}}

\newcommand{\bP}[0]{\mathbf{P}}

\newcommand{\br}[0]{\mathbf{r}}
\newcommand{\bR}[0]{\mathbf{R}}
\newcommand{\bs}[0]{\mathbf{s}}

\newcommand{\bV}[0]{\mathbf{V}}

\newcommand{\by}[0]{\mathbf{y}}

\newcommand{\bZ}[0]{\mathbf{Z}}

\newcommand{\bthetat}[0]{\boldsymbol{\tilde{\theta}}}

\newcommand{\brh}[0]{\mathbf{\hat{r}}}

\newcommand{\bRh}[0]{\mathbf{\hat{R}}}

\newcommand{\bthetah}[0]{\boldsymbol{\hat{\theta}}}

\newcommand{\zeros}[0]{\mathbf{0}}

\newcommand{\MCE}[0]{\mathcal{E}}

\newcommand{\expect}[1]{\MCE\{#1\}}

\newcommand{\beq}{\begin{equation}}
\newcommand{\eeq}{\end{equation}}
\newcommand{\bea}{\begin{array}}
	\newcommand{\ena}{\end{array}}

\newcommand{\DL}{\begin{dashlist}}
	\newcommand{\DLE}{\end{dashlist}}